\documentclass[12pt]{article}
\usepackage{amsfonts}
\usepackage{graphicx}
\usepackage{amsmath}
\usepackage{float}
\textwidth=6.5in \hoffset=-.75in \voffset=-1in \numberwithin{equation}{section}
\numberwithin{figure}{section}

\newcommand {\nn}{\nonumber}
\newcommand {\be}{\begin{equation}}
\newcommand {\ee}{\end{equation}}
\begin{document}

\begin{titlepage}
\vspace{1cm}
\begin{center}
{\Large \bf { Cosmological solutions in five dimensional Einstein-Maxwell-dilaton theory}}\\
\end{center}
\vspace{2cm}
\begin{center}
A. M. Ghezelbash{ \footnote{ E-Mail: masoud.ghezelbash@usask.ca}}
\\
Department of Physics and Engineering Physics, \\ University of Saskatchewan, \\
Saskatoon, Saskatchewan S7N 5E2, Canada\\
\vspace{1cm}
\vspace{2cm}
%\today\\
\end{center}

\begin{abstract}
We construct new classes of exact cosmological solutions to five dimensional Einstein-Maxwell-dilaton theory with two coupling constants for the dilaton-Maxwell term and dilaton-cosmological constant term. All the solutions are non-stationary and the solutions that both coupling constants are non-zero are almost regular everywhere. The size of spatial section of the asymptotic metric shrinks to zero at early time and increases to infinitely large at very late time. The cosmological constant depends on the dilaton coupling constant and can take positive, zero or negative values. 

%The solutions are regular everywhere and show a bolt structure on a single point in any dimensionality.
%Moreover, we find the exact non-stationary solutions to the Einstein-Maxwell theory with positive cosmological constant. We show that the cosmological solutions are expanding patches in asymptotically de Sitter spacetime.

\end{abstract}
\end{titlepage}\onecolumn 
\bigskip 

\section{Introduction}
The exact solutions to the Einstein gravity in presence of matter fields are the building blocks to explore and understand the realm of the gravitational physics in any dimensionalities. The asymptotically locally flat, dS and AdS solutions to Einstein-Maxwell theory in four and higher dimensions with NUT charges have been found in \cite{awad}. Moreover, the black hole solutions with different non-trivial horizons in five dimensions have been found in \cite{Myers1}-\cite{Jap}. Including the dilaton field as well as axion, as the simplest matter fields to the Einstein-Maxwell theory, opens the door to new solutions and their physical properties in the Einstein-Maxwell-dilaton-(axion) theory with/without the cosmological constant and the Chern-Simons term \cite{biglist1}-\cite{biglist21}. Some other interesting solutions such as supergravity solutions, as well as solitonic and dyonic solutions have been found in \cite{biglist3}-\cite{biglist32}.

Moreover in the context of generalized Freund-Rubin compactification with cosmological constant and the dilaton field, the Einstein-Maxwell-dilaton theory with two different coupling constants for dilaton-Maxwell term and dilaton-cosmological constant term has been considered in \cite{Torii}-\cite{Kino}. The cosmological solutions to the Einstein-Maxwell-dilaton theory with two coupling constants were found in \cite{Mak} in which the spatial section of the metric is Euclidean space.
Motivated by the cosmological application of these solutions to the very early evolution of the universe, in this article we find new class of exact solutions to the Einstein-Maxwell-dilaton theory with two coupling constants in which the spatial section of the solutions contains a NUT charge.
More specifically, we consider different possibilities for two dilaton coupling constants and find exact analytical solutions to the equations of motion. In all cases, the spacetime metric is non-stationary. We show that in the special case where the coupling constant for the dilaton-cosmological constant term is not zero, while the other coupling constant is zero, we have an exact solution for the spacetime where the scale factor and the dilaton field depend only on time. 

The article is organized as follows. In section \ref{sec:ab}, we consider the Einstein-Maxwell-dilaton theory in presence of cosmological constant in which the dilaton coupling constant to the Maxwell field is different from coupling constant to the cosmological constant. We employ an ansatz for the metric in which the spatial section of the metric has separable metric functions in time and the radial coordinate. We solve the equations of motion and find that the two dilaton coupling constants are related by a simple relation. Moreover, we find that the cosmological constant is related to one of the dilaton coupling constant and can be positive, zero or negative. We discuss the asymptotic of the metric and behaviours of dilaton and Maxwell field strength. In section \ref{sec:aa}, we consider the Einstein-Maxwell-dilaton theory in which the two dilaton coupling constants are equal. We use a different metric ansatz that resembles the metric ansatz in section \ref{sec:ab}, however one of the metric functions depends explicitly on both time and radial coordinate. The form of Maxwell field is also different from the form of Maxwell field in section \ref{sec:ab}. We then solve all the equations of motion and find explicit exact solutions for the metric functions and the dilaton filed. We find specific values for the cosmological constant in terms of dilaton coupling constant and discuss the physical properties of the solutions. Moreover, we consider the special case in which the dilaton coupling constant is zero and so the theory effectively reduces to the Einstein-Maxwell theory. We wrap up the article by concluding remarks in \ref{sec:con}.

\section{The Einstein-Maxwell-dilaton theory with $a\neq b$}
\label{sec:ab}
We consider the Einstein-Maxwell-dilaton theory in five dimensions, which the dilaton field couples to the Maxwell field as well as the cosmological constant. The action is given by
\begin{equation}
S=\frac{1}{16\pi}\int d^5x \sqrt{-g}\{ R-\frac{4}{3}(\nabla \phi)^2-e^{-4/3a\phi}F^2-e^{4/3b\phi}\Lambda\},
\label{act}
\end{equation}
where $a$ and $b$ are the coupling constants for coupling of the dilaton to the Maxwell field strength and the cosmological constant, respectively. The most interesting case of theory with $a\neq b\neq 0$ in different dimensionalities has been considered in the context of generalized Freund-Rubin compactification with cosmological constant and the dilaton field \cite{Torii}-\cite{Kino}.  Moreover, we note that the action for the Einstein-Maxwell theory is the special case of action (\ref{act}) with $a=b=0$.

Varying the action (\ref{act}) with respect to the metric tensor yields the Einstein's equations 
\begin{equation}
{\cal G}_{\mu\nu}=R_{\mu\nu}-\frac{2}{3}\Lambda g_{\mu\nu}e^{4/3b\phi}-(F_{\mu}^{\lambda}F_{\nu\lambda}-\frac{1}{6}g_{\mu\nu}F^2)e^{-4/3a\phi}-\frac{4}{3}\nabla_\mu \phi \nabla _\nu \phi=0,\label{eins}
\end{equation}
while the equations of motion for the Maxwell field and dilaton field are 
\begin{eqnarray}
{\cal F}^\nu=\nabla _\mu (e^{-4/3a\phi}F^{\mu\nu})&=&0,\label{eqf}\\
\nabla ^2 \phi + a/2e^{-4a\phi/3}F^2-b/2e^{4/3b\phi}\Lambda&=&0\label{eqphi}.
\end{eqnarray}
In this section, we consider the most general case of the action (\ref{act}) where both $a$ and $b$ are non-zero and seek the five-dimensional solution of the form
\begin{equation}
ds_5^{2}=-\frac{1}{H(r)^{2}}dt^{2}+R(t)^2H(r)ds_{n}^2,
\label{ds6}
\end{equation}
where $ds_{n}^2$ represents the four-dimensional multi-center Taub-NUT (TN) space. To simplify the calculation, we consider the one-center TN space that is given by
\begin{equation}
ds_{n}^{2}=V(r)(dr^{2}+r^2d\Omega^2) +\frac{(d\psi+n\cos\theta d\phi)^2}{V(r)}.\label{Nmetric}
\end{equation}%
The function $V(r)$ in (\ref{Nmetric}) is given by $V(r)=1+\frac{n}{r}$ in terms of a positive NUT charge $n$. We note that the periodic coordinate $\psi$ parameterizes the fibration of a circle over the sphere $(\theta,\phi)$ and is restricted to the interval $[0,4\pi n]$. 
We only consider a non-vanishing temporal component for the Maxwell field as
\be
{A_t}(t,r)=\alpha R(t)^2({F(r)}-\beta)\label{gauge},
\ee
that leads to a radial electric field, where $\alpha$ and $\beta$ are two constants and $F(r)$ is a function of coordinate $r\geq 0$.
Moreover, we consider the dependence of the dilaton field to the metric functions $R(t)$ and $H(r)$ as
\be
\phi (t,r)=-\frac{3}{4a}\ln \{ R(t)^\gamma H(r)^{\delta} \},\label{dila}
\ee
where $\gamma$ and $\delta$ are two other constants. 
The only non-zero components of ${\cal F}$ in equation (\ref{eqf}) are ${\cal F}^t$ and ${\cal F} ^r$. The first component determines the function $F(r)$ in terms of metric function $H(r)$ as 
\be 
F(r)=F_1+h\int \frac{dr}{H(r)^{\delta +2}r^2}\label{FvH},
\ee
for any two constants $F_1$ and $h$. The second component ${\cal F} ^r$ implies $\gamma=-4$. The equation (\ref{eins}) for ${\cal G}_{tr}$ fixes the constant $\delta$ to be equal to $a^2$. So, we find out that the dilaton field is
\be
\phi (t,r)=-\frac{3}{4a}\ln \{\frac{H(r)^{a^2}}{R(t)^4} \}\label{dilfinal}.
\ee
The other remaining equations of motion (Appendix A) lead to the results for the metric functions
\be
H(r)=(\frac{r+h}{r})^{\frac{2}{2+a^2}},\,R(t)=R_0t^{\frac{a^2}{4}},\label{HandRcase3}
\ee
where $R_0$ is a constant and we set it equal to one in what follows.
From equation (\ref{FvH}), we then find $F(r)=\frac{r}{r+h}$. Moreover, the equations of motion imply that the dilaton coupling constant $b$ is $b=-\frac{2}{a}$. The constant $\alpha$ in (\ref{gauge}) is related to the dilaton coupling constant $a$ by $\alpha^2=\frac{3}{2+a^2}$ and the cosmological constant is equal to
\be
\Lambda=\frac{3a^2}{8}(a^2-1).\label{Lam3}
\ee
As we notice, the cosmological constant can be positive, zero or negative depending on the value of the dilaton coupling constant $a$. Furnished with all metric functions, the
spacetime metric (\ref{ds6}) then read as
\be
ds_5^2=-(\frac{r}{r+h})^{\frac{4}{2+a^2}}dt^{2}+t^{\frac{a^2}{2}}(\frac{r+h}{r})^{\frac{2}{2+a^2}}V(r)\{dr^{2}+r^2d\Omega^2 +\frac{(d\psi+n\cos\theta d\phi)^2}{V(r)^2}\}.\label{case3metr}
\ee
In asymptotic region $r\rightarrow \infty$, the metric (\ref{case3metr}) reduces to
\be
ds_5^2=-dt^{2}+t^{\frac{a^2}{2}}\{dr^{2}+r^2d\Omega^2 +(d\psi+n\cos\theta d\phi)^2\},\label{case3metrasym}
\ee
which for a fixed $t$ slice, represents the fibration of a circle over the $S^2$. The Ricci scalar of the asymptotic metric (\ref{case3metrasym}) is divergent at $t=0$ due to vanishing of spatial section of the asymptotic metric. However, if we rescale the asymptotic metric by the asymptotic value of a conformal factor $\lim _{r \rightarrow \infty} e^{-4/3a\phi}$, the Ricci scalar approaches to $t^{a^2 -2}(\frac{1}{4}a^4+2a^2)$ for $r\rightarrow \infty$. Hence by restricting the dilaton coupling $a\geq 2$, the rescaled asymptotic spacetime has zero or a constant positive Ricci scalar. In fact, 
both the Ricci scalar and the Kretschmann invariant for the metric (\ref{case3metr}) diverge for small values of the radial coordinate as well as time. However if we rescale the metric (\ref{case3metr}) by the conformal factor $e^{-4/3a\phi}$, we find that the Ricci scalar and the Kretschmann invariant of the rescaled metric
are finite for $r\rightarrow 0$ as well as $t\rightarrow 0$ as long as the dilaton coupling constant $a\geq 2$.

\begin{figure}[H]%HAVE BOTH .eps and converted to PDF in DIRECTORY
\centering
\includegraphics[width=0.5\textwidth]{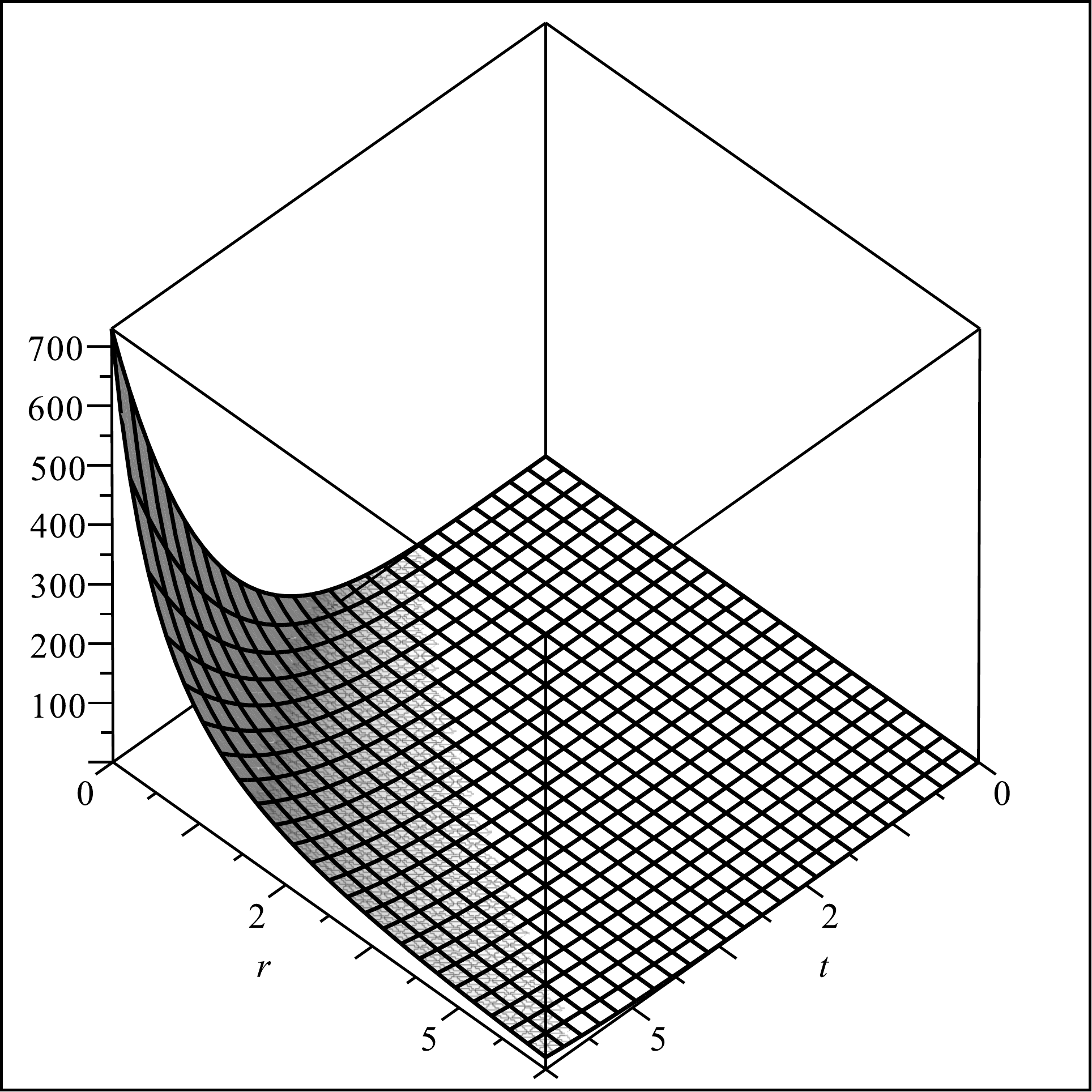},\includegraphics[width=0.5\textwidth]{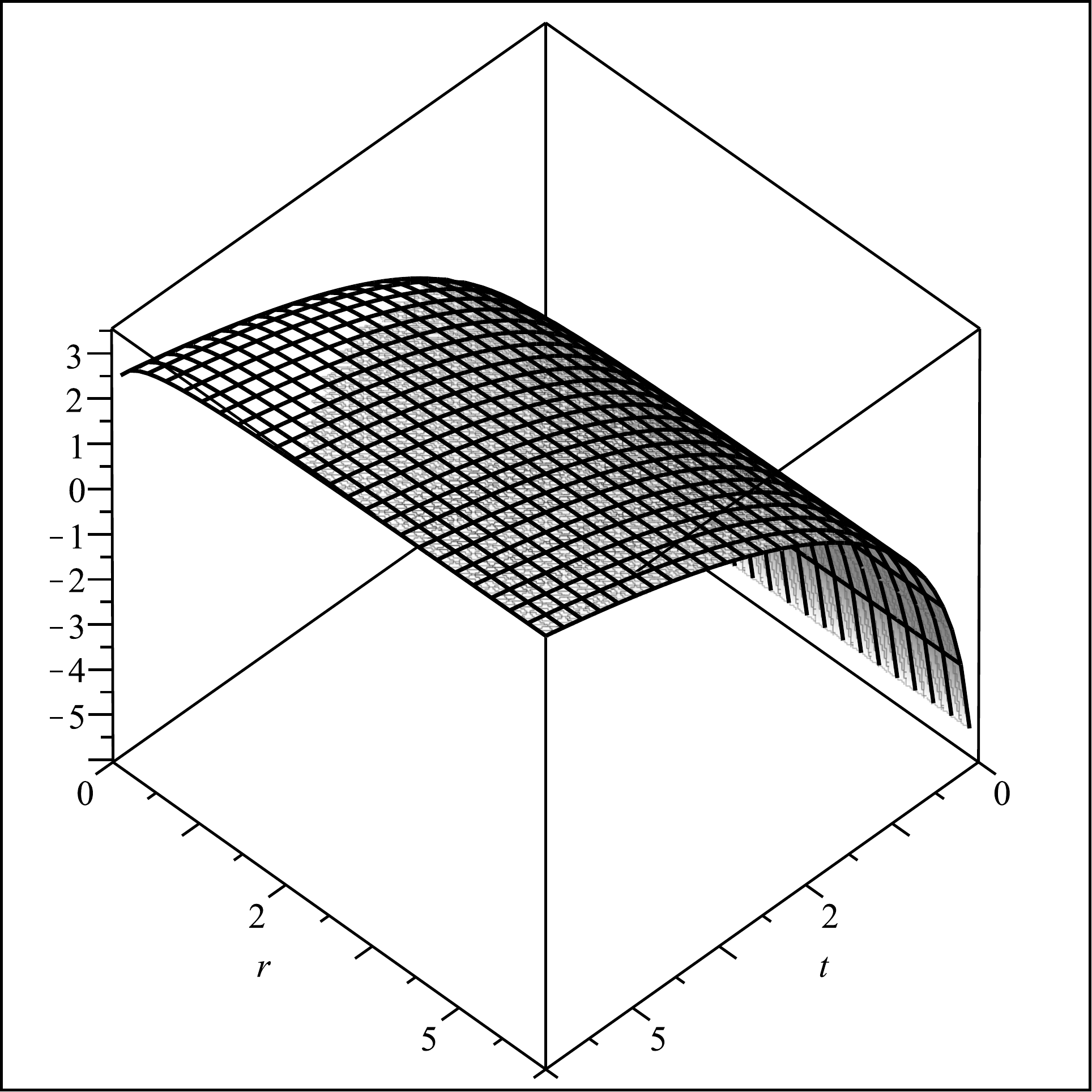}
% \includegraphics[width=0.5\textwidth]{Kummer1}
% f1.png: 0x30 pixel, 0dpi, 0.00xinf cm, bb=
\caption{The magnitude of $F_{tr}$ (left) and dilaton (right ) as function of radial coordinate and time.}
\label{fig1}
\end{figure}

The only non-zero component of the Maxwell field strength is $F_{t,r}=\sqrt{\frac{3}{2+a^2}}\frac{h}{(r+h)^2}t^{\frac{a^2}{2}}$. Figure \ref{fig1} shows the behaviour of the Maxwell field strength and the dilaton field for small values of radial coordinate $r$ and time $t$ where we set $a=3$ and $h=1$. 
We may expect for the special value of the dilaton coupling constant $b\rightarrow 0$ which means $a\rightarrow -\infty$, we recover the traditional asymptotically dS results in Einstein-Maxwell-dilaton theory. However, the limit of the solution (\ref{case3metr}) where $a\rightarrow -\infty$ and $b\rightarrow 0$ doesn't correspond to any solution in asymptotically dS spacteimes. In fact, in these limits, we find that the Ricci scalar for the solution (\ref{case3metr}), is time dependent of the form $\frac{5a^4}{4t^2}$ while the cosmological constant is $\Lambda=\frac{3a^4}{8}$ in contrast to the well known relation between the Ricci scalar and the cosmological constant of the asymptotically dS spacetimes. However in the next section, we consider the solutions for the theory (\ref{act}) with $a=b$ and recover the known solutions in the traditional Einstein-Maxwell theory in the limit where $a=b\rightarrow 0$. 

%\begin{figure}[H]%HAVE BOTH .eps and converted to PDF in DIRECTORY
%\centering
% \includegraphics[width=0.5\textwidth]{Kummer1}
% f1.png: 0x30 pixel, 0dpi, 0.00xinf cm, bb=
%\caption{The dilaton as a function of radial coordinate and time}
%\label{fig2}
%\end{figure}

\section {The Einstein-Maxwell-dilaton theory with $a=b$}
\label{sec:aa}

We consider the action (\ref{act}) with $a=b$ and seek a solution of the form 
\begin{equation}
ds_5^{2}=-\frac{1}{H(t,r)^{2}}dt^{2}+R(t)^2H(t,r)ds_{n}^2,
\label{ds6aeqb}
\end{equation}
along with the Maxwell gauge field
\be
{A_t}(t,r)=\frac{\alpha}{ R(t)^{a^2}}({F(t,r)}-\beta)\label{gaugeaeqb}.
\ee
We note that the metric function $H$ in (\ref{ds6aeqb}) depends on both the time coordinate $t$ and the radial coordinate $r$ unlike the dependence of metric function in (\ref{ds6}) on the radial coordinate only.
We consider the dilaton field as given by (\ref{dila}). The equation $ {\cal F}^t$ provides the solution for $F(t,r)$ as
\be
F(t,r)=F_1(t)+F_2(t)\int \frac{dr}{r^2H(t,r)^{2+\delta}}.
\ee
The second component of Maxwell's equation ${\cal F}^r$ along with the components of ${\cal G}$ imply $\gamma=2a^2$, $\delta=a^2$ and $F_2(t)=hR(t)^{-2-a^2}$, hence we find
\be
\phi (t,r)=-\frac{3}{4a}\ln \{R(t)^{2a^2}H(t,r)^{a^2}\}.\label{phicase2}
\ee
Moreover, the other remaining equations of motion lead to the metric function
\be
H(t,r)=\{1+\frac{h}{rR(t)^{2+a^2}}\}^{\frac{2}{2+a^2}},
\ee
where the function $R(t)$ turns out to be $R_0t^{\frac{1}{a^2}}$ and so $F(t,r)=\frac{rR(t)^{2+a^2}}{h+rR(t)^{2+a^2}}$ where $R_0$ and $h$ are constants and we chose $F_1(t)=0$.
Moreover, we find that the cosmological constant is related to the dilaton coupling constant by 
\be
\Lambda=\frac{3(4-a^2)}{2a^4},
\ee
and we also find $\alpha^2=\frac{3}{2+a^2}$. We note that depending on the dilaton coupling constant $a$, the cosmological constant takes positive and negative values as well as zero. 
Also, we notice that the only non-zero component of the Maxwell field strength is $F_{tr}$, that is given by 
\be
F_{tr}=\alpha\,h \frac{1}{\left( tr+{t}^{-2\,{a}^{-2}}h \right)\left( {t}^{{
\frac {{a}^{2}+2}{{a}^{2}}}}r+h \right)}.
\ee

\begin{figure}[H]%HAVE BOTH .eps and converted to PDF in DIRECTORY
\centering
\includegraphics[width=0.5\textwidth]{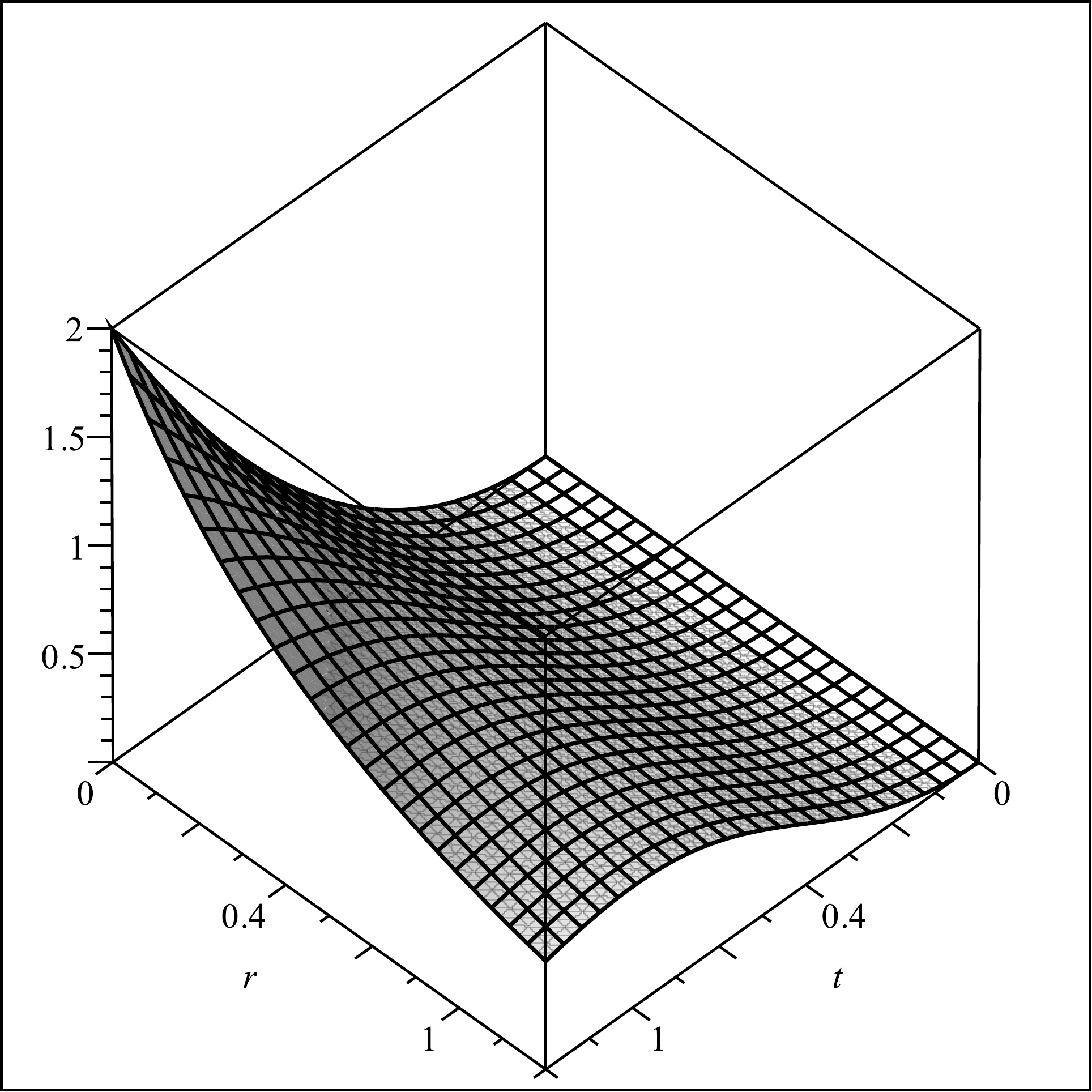},\includegraphics[width=0.5\textwidth]{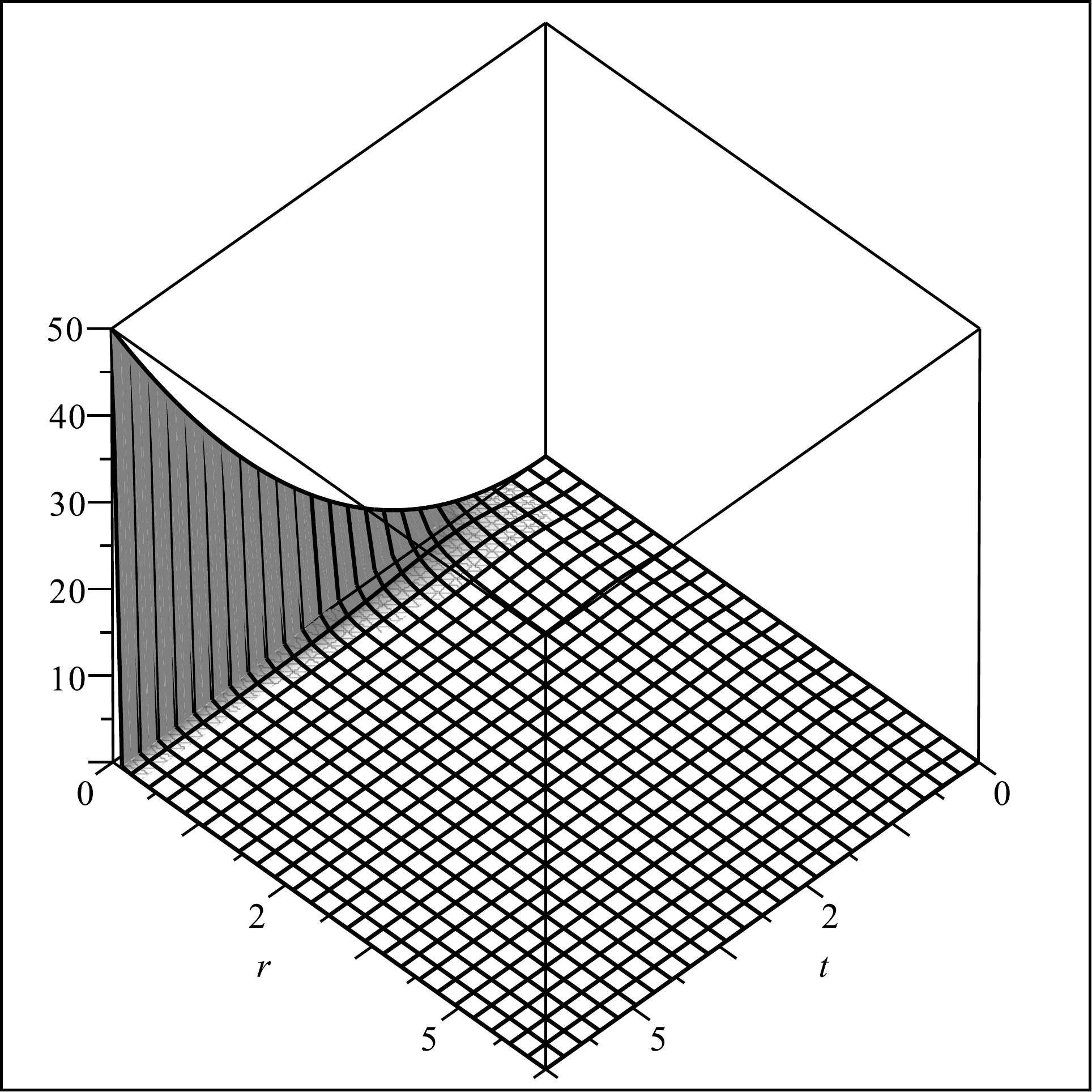}
% \includegraphics[width=0.5\textwidth]{Kummer1}
% f1.png: 0x30 pixel, 0dpi, 0.00xinf cm, bb=
\caption{The magnitude of $F_{tr}$ as a function of radial coordinate and time.}
\label{fig3}
\end{figure}

\begin{figure}[H]%HAVE BOTH .eps and converted to PDF in DIRECTORY
\centering
\includegraphics[width=0.5\textwidth]{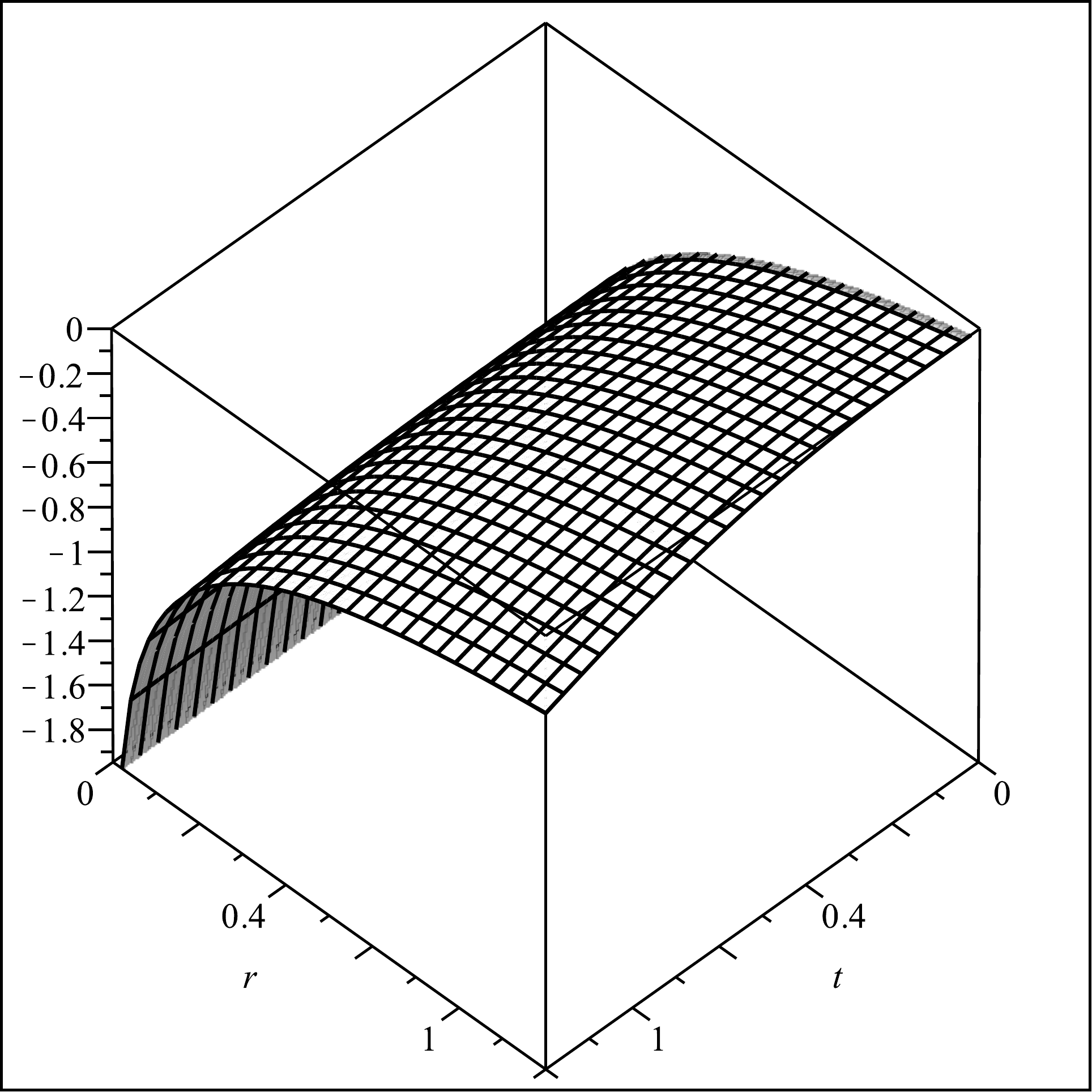},\includegraphics[width=0.5\textwidth]{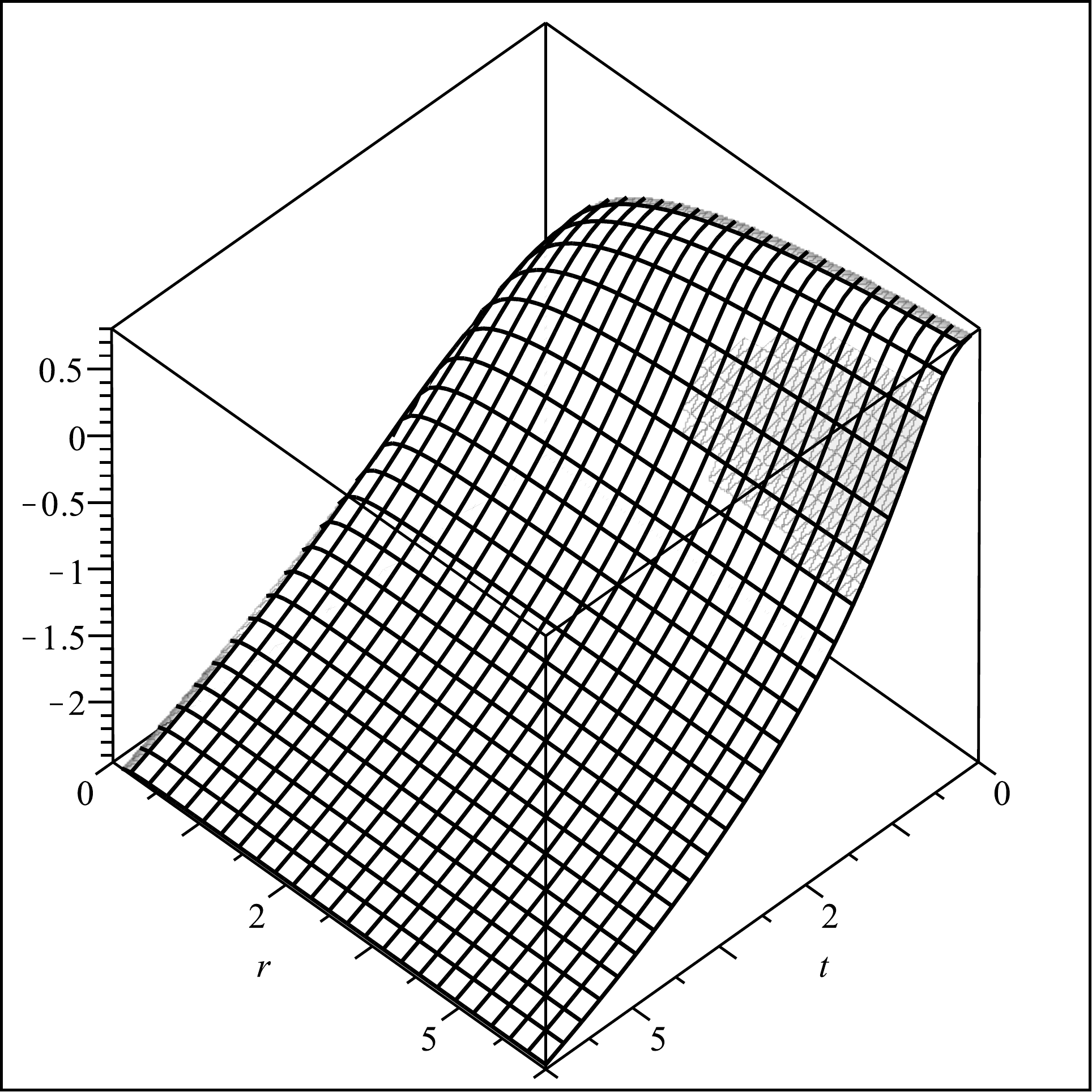}
% \includegraphics[width=0.5\textwidth]{Kummer1}
% f1.png: 0x30 pixel, 0dpi, 0.00xinf cm, bb=
\caption{The dilaton as a function of radial coordinate and time.}
\label{fig4}
\end{figure}

Figures \ref{fig3} and \ref{fig4} show respectively the typical behaviour of $F_{tr}$ and the dilaton field as function of time and radial coordinate where we set $ h = 1, \alpha = 2, a = 1$.\newpage

\newpage
So, furnished by the solutions for $H(t,r)$ and $R(t)$, the metric (\ref{ds6aeqb}) is given by 

\be
ds^2=t^{\frac{4}{a^2}}r^{\frac{4}{2+a^2}}(h+rt^{\frac{2+a^2}{a^2}})^{\frac{-4}{2+a^2}}dt^2+\frac{r+n}{r^{\frac{4+a^2}{2+a^2}}}(h+rt^{\frac{2+a^2}{a^2}})^{\frac{2}{2+a^2}}\{dr^{2}+r^2d\Omega^2 +\frac{(d\psi+n\cos\theta d\phi)^2}{V(r)^2}\}.\label{metraeqb}
\ee
To explore the properties of the metric (\ref{metraeqb}), we first consider the asymptotic limit $r\rightarrow \infty$ where the metric reduces to
\be
-dt^2+t^{2/a^2}\{dr^{2}+r^2d\Omega^2 +{(d\psi+n\cos\theta d\phi)^2}\}.\label{case2asym}
\ee
The equal-time hypersurfaces $t=\text{constant}$ represent the fibration of a circle over the $S^2$. The scale factor for this hypersurface is $t^{2/a^2}$, while the scale factor for the asymptotic limit of solutions with $a\neq b$ is $t^{a^2/2}$. Moreover, we find that the Ricci scalar of the asymptotic metric (\ref{case2asym}) is divergent at $t=0$ due to vanishing of spatial section of the asymptotic metric. The size of spatial section of the asymptotic metric increases to infinitely large as $t\rightarrow +\infty$.
The Ricci scalar of the full metric (\ref{metraeqb}) is divergent at $r=0$ for $a\geq 2$. The Kretschmann invariant is divergent for any value of $a$ and so $r=0$ is the location of a bolt structure in the spacetime. We note that we can't simply take the limit $a\rightarrow 0$ to understand the behaviour of our solutions where the dilaton coupling constant approaches zero. As an example, in this limit, the dilaton field (\ref{phicase2}) doesn't approach to any well-defined function. So, to understand this limiting case, we consider the action (\ref{act}) with $a=b=0$. The equation of motion (\ref{eqphi}) implies that the dilaton decouples from the theory and so the action reduces to that of the Einstein-Maxwell theory with a cosmological constant. We consider the five-dimensional metric in the form of (\ref{ds6aeqb}) and the Maxwell field given by $A_t(t,r)=\alpha(F(t,r)-\beta)$. We then find that the equations of motion imply the metric function is given by
\be
H(t,r)=1+\frac{h}{R(t)^2r},\label{Habzero}
\ee
where
\be
R(t)=R_0e^{\gamma t}\label{Rabzero}.
\ee
Moreover, we find that $F(t,r)=\frac{1}{H(t,r)}$ and $\alpha^2=\frac{3}{2}$. The cosmological constant $\Lambda=6\gamma ^2$. This special solution is in agreement with the solution that was previously obtained in \cite{Jap}. The spacetime unlike the other cases (\ref{case3metr}) and (\ref{metraeqb}) is asymptotically dS. In fact, the metric of spacteime for $r\rightarrow \infty$ has the same form as (\ref{case3metrasym}) or (\ref{case2asym}) with the scale factor of $e^{\gamma t}$ for the spatial section of the asymptotic metric. The equal-time hypersurfaces $t=\text{constant}$ represent the fibration of a circle over the $S^2$, similar to the former cases with non-zero dilaton coupling constants $a$ and $b$. However the hypersurfaces are expanding (for $\gamma=+\sqrt{\frac{\Lambda}{6}}$) or shrinking (for $\gamma=-\sqrt{\frac{\Lambda}{6}}$) patches of dS, while for the former cases (with non-zero dilaton coupling constants) are expanding patches of a non-static spacetime by scale factors of $t^{a^2/2}$ and $t^{2/a^2}$, respectively. In fact, the proposed c-function for the five-dimensioanl asymptotically dS spacetimes \cite{Lob}, $c \sim \frac{1}{G_{\mu\nu}^{3/2}}$ for the solution ({\ref{ds6aeqb}) with (\ref{Habzero}) and (\ref{Rabzero}) shows that the the renormalization group flows toward the ultraviolet or infrared depending on the value of $\gamma >0$ or $\gamma <0$. %or ultraviolet depensolution is asymptotically locally dS is indeed made of patch . 
Figure \ref{fig5} shows the c-functions for $\gamma >0$ or $\gamma <0$ where we set $h = 1, \Lambda = 6, r = 1, n = 2$.

\begin{figure}[H]%HAVE BOTH .eps and converted to PDF in DIRECTORY
\centering
\includegraphics[width=0.5\textwidth]{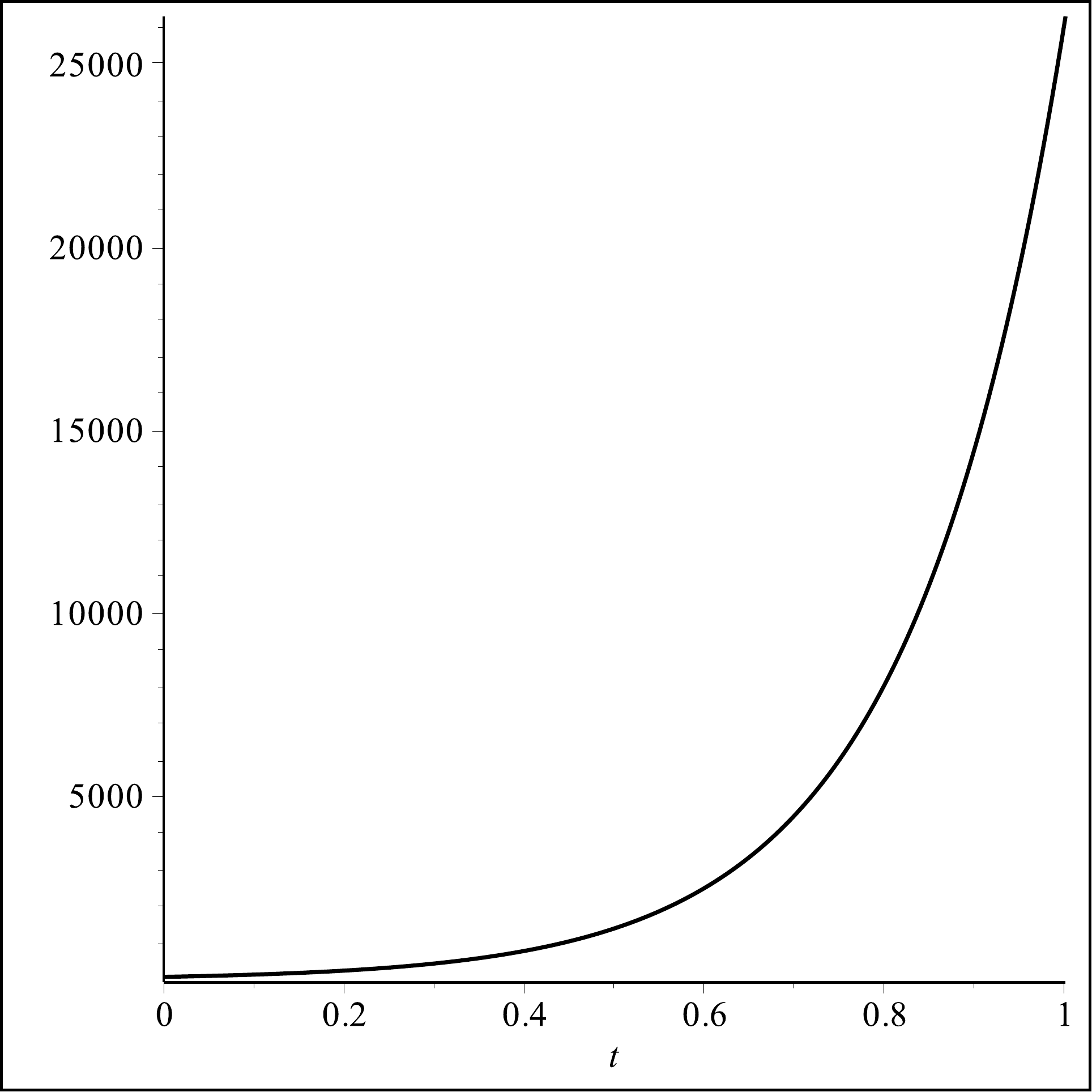},\includegraphics[width=0.5\textwidth]{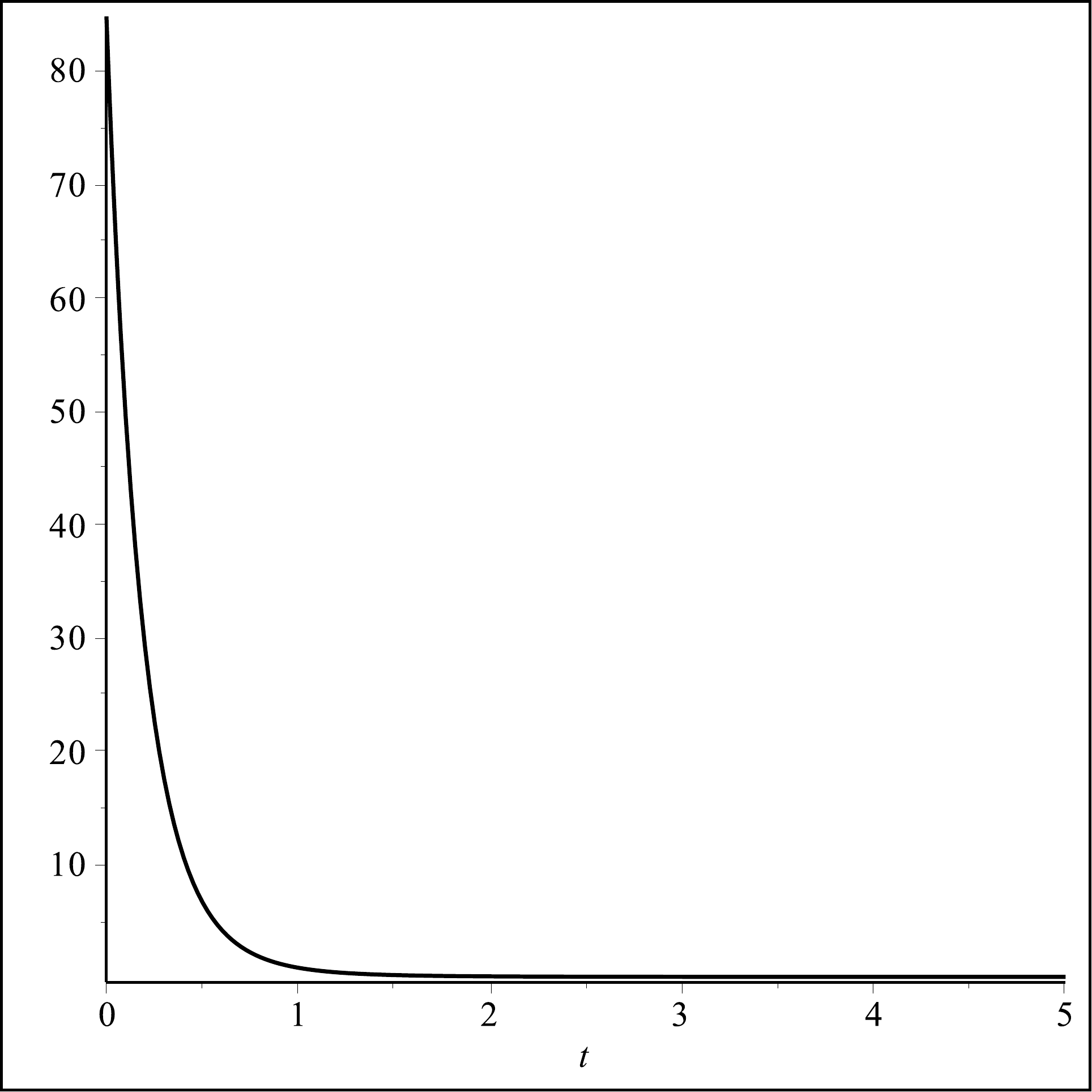}
% \includegraphics[width=0.5\textwidth]{Kummer1}
% f1.png: 0x30 pixel, 0dpi, 0.00xinf cm, bb=
\caption{The c-functions for the cosmological solutions with $a=b=0$. (left) $\gamma=\sqrt{\frac{\Lambda}{6}}$ and (right) $\gamma=-\sqrt{\frac{\Lambda}{6}}$.}
\label{fig5}
\end{figure}

To understand the meaning of the dilaton coupling constant $b\neq 0$ in the solutions, we set $a=0$ and moreover consider no Maxwell fields in the action (\ref{act}), so the theory reduces simply to the Einstein-dilaton gravity in presence of cosmological constant. We seek the solutions for the metric given by
\be
ds^2=-dt^2+R(t)^2ds_n^2
\ee
Inspired with previous solutions in which the dilaton coupling constants are not zero, we consider the metric function $R(t)=R_0 t^A$ and the dilaton field as $\phi(t)=\frac{3}{4b}\ln t^B$. All equations of motion are satisfied with $A=\frac{1}{b^2}$ and $B=-2$ and the cosmological constant
\be
\Lambda=\frac{3}{2}{\frac {4-{b}^{2}}{{b}^{4}}}
\ee
We notice the cosmological constant is positive for the dilaton coupling $b<2$ and negative for $b>2$.

\section{Concluding Remarks}
In this article, we construct exact cosmological solutions to Einstein-Maxwell-dilaton theory in five dimensions with different values for two dilaton coupling constants. All solutions are non-stationary where the dilaton field and the Maxwell field depend on the radial and time dependent metric functions. In the first class of solutions, the two coupling constants are different and the scale factor for the spatial section of the metric separates in the radial and time coordinates. The solution is regular almost everywhere except on the location of NUT charge. The cosmological constant depends on dilaton coupling constant and can take positive, zero or negative values. We discuss in detail the properties of the asymptotic metric as well as the behaviour of the Maxwell field strength and dilaton field. In the second class of solutions, we consider equal dilaton coupling constants and find the metric, dilaton and the Maxwell fields. Finally, in the special case which both dilaton coupling constants are zero, the solutions reduce to the results obtained in \cite{Jap}. We find the c-function for the latter solutions and show that the flow of renormalization group is in agreement with the c-theorem for an asymptotically dS spacetime. We note that the generalization to multi-center TN is straightforward where both dilaton coupling constants are zero. However, it is an interesting task to find the exact solutions where the dilaton coupling constants are not zero. Moreover, it is interesting to find the exact solutions in five or higher dimensions where the spatial section is any other four dimensional geometries as well as higher dimensional TN spaces. Some of these higher dimensional solutions can be convoluted-like solutions \cite{Conv}. The other interesting case is to find the exact solutions for Einstein-Maxwell-dilaton theory with Chern-Simons term. We leave these open questions as well as investigating the thermodynamics of the present solutions for a future article.

\label{sec:con}

\bigskip
\section{Appendix A}
We use equations (\ref{FvH}) and (\ref{dilfinal}) to eliminate $F(r)$ and $\phi(t,r)$ from the equations of motion (\ref{eins}) and (\ref{eqphi}).  We find the numerators of ${\cal G}_{\mu\nu}$ in (\ref{eins}) (that we denote by $\bar {\cal G}_{\mu\nu}$) are given by
%the numerators given byThe other equations of motion are 
%THEY ARE ALL NUMERATORS OF G?? SO CORRECT THEM 
\begin{eqnarray}
\bar{\cal G}_{tt}&=&-2{\alpha}^{2} h^2 H \left( r \right)^{-2\,{a}^{2}}
 R \left( t \right) ^{4}{a}^{2}-36V \left( r
 \right)  H \left( r \right)  ^{-{a}^{2}+5} R
 \left( t \right) ^{4}{r}^{4} \left( {\frac {d}{dt}}R \left( 
t \right)  \right) ^{2}\nn \\&+&2\Lambda\, R \left( t \right) 
^{2\,{\frac {2\,b+3\,a}{a}}} H \left( r \right) 
 ^{-{a}^{2}-ab+3}V \left( r \right) {r}^{4}{a}^{2}-12H
 \left( r \right) ^{-{a}^{2}+5} R \left( t \right) 
 ^{5}V \left( r \right) {r}^{4}{a}^{2}
 \left( {\frac {d^{2}}{d{t}^{2}}}R \left( t \right)  \right)  \nn \\&-&3  H \left( r \right) 
 ^{- \left( a-1 \right)  \left( a+1 \right) } R \left( 
t \right) ^{4}{r}^{4}{a}^{2}{\frac {d^{2
}}{d{r}^{2}}}H \left( r \right)   +3R \left( t \right) ^{4
} \left( H \left( r \right)  \right) ^{-{a}^{2}}{r}^{4}{a}^{2}\left( {\frac {d}{dr}}H
 \left( r \right)  \right) ^{2} \nn \\&-&6
 H \left( r \right)  ^{- \left( a-1 \right)  \left( a+1
 \right) }  R
 \left( t \right)   ^{4}{r}^{3}{a}^{2}{\frac {d}{dr}}H \left( r \right) ,\label{E1}
\end{eqnarray}
\begin{eqnarray}
\bar{\cal G}_{rr}&=&12 
  R \left( t \right)  ^{5}  H \left( r \right) 
 ^{-{a}^{2}+5} V \left( r \right)  ^{2}{r}^{4}{\frac {d^{2}}{d{t}^{2}}}R \left( t \right)+
36\,  R \left( t \right)  ^{4} H \left( r
 \right)  ^{-{a}^{2}+5} V \left( r \right)  ^{2
} r^4\left( {\frac {d}{dt}}R \left( t \right)  \right) ^{2}\nn\\&-&6
R
 \left( t \right) ^{4} H \left( r \right) ^{-
{a}^{2}+1}V \left( r \right) {r}^{4} {\frac {d^{2}}{d{r}^{2}}}H \left( r \right) -6R \left( t \right) ^{4
} H \left( r \right) ^{-{a}^{2}+2}{r}^{4}{\frac {d^{2}}{d{r}^{2
}}}V \left( r \right) \nn\\&-&12
R \left( t \right) ^{4}  H \left( r \right) ^{
-{a}^{2}}V \left( r \right)  r^4\left( {\frac {d}{dr}}H \left( r \right) 
 \right) ^{2}-12R \left( t \right) ^{4}
 H \left( r  \right) ^{-{a}^{2}+1}V \left( r \right) {r}^{3}
 {\frac {d}{dr}}H \left( r \right) \nn\\&-&12
  R \left( t   \right) ^{4}H \left( r \right) 
  ^{-{a}^{2}+2} {r}^{3}{\frac {d}{dr}}V \left( r 
 \right) -8\Lambda R \left( t \right) ^{2\,
{\frac {2\,b+3\,a}{a}}} H \left( r \right) ^{-{a}^{2}-
ab+3} V \left( r\right) ^{2}{r}^{4}\nn\\&+&8\,{\alpha}^{2}h^2
 H \left( r   \right) ^{-2\,{a}^{2}} R \left( t
 \right) ^{4}V \left( r \right) -9\,{r}^{4}V \left( r
 \right) R \left( t \right) ^{4} H \left( r
 \right) ^{-{a}^{2}}{a}^{2} \left( {\frac {d}{dr}}H \left( r
 \right)  \right) ^{2},\label{E2}
\end{eqnarray}

\begin{eqnarray}
\bar{\cal G}_{\theta\theta}&=&
6\,  {\frac {d^{2}}{d{t}^{2}}}R \left( t \right)   
  R \left( t \right) ^{5}  H \left( r \right) 
 ^{-{a}^{2}+5} V \left( r  \right)^{3}{r}^{4}+
18\, R \left( t  \right) ^{4} H \left( r \right) ^{-{a}^{2}+5} V \left( r \right) ^{3
}{r}^{4} \left( {\frac {d}{dt}}R \left( t \right)  \right) ^{2}\nn\\&-&3
R
 \left( t \right) ^{4} H \left( r  \right) ^{-
{a}^{2}+1} V \left( r \right) ^{2}{r}^{4} {\frac {d^{2}}{d{r}^{2}}}H \left( r \right)  -3 R \left( t
 \right) ^{4} H \left( r  \right) ^{-{a}^{2}+2
}V \left( r \right) {r}^{4}{
\frac {d^{2}}{d{r}^{2}}}V \left( r  \right) \nn\\&+&3R \left( t  \right) ^{4
} H \left( r \right)   ^{-{a}^{2}} V \left( r
 \right)  ^{2}{r}^{4} \left( {\frac {d}{dr}}H \left( r \right) 
 \right) ^{2}+3R \left( t \right) ^{4}
 H \left( r  \right) ^{-{a}^{2}+2} {r}^{4}\left( {\frac {d}{dr
}}V \left( r \right)  \right) ^{2}\nn\\&-&6R \left( t
 \right)  ^{4} H \left( r \right)  ^{-{a}^{2}+1
} V \left( r \right) ^{2} {r}^{3}{\frac {d}{dr}}H
 \left( r   \right) -6 R \left( t \right) 
  ^{4} H \left( r \right) ^{-{a}^{2}+2}V
 \left( r \right) {
r}^{3} {\frac {d}{dr}}V \left( r  \right) \nn\\&-&3R \left( t  \right) ^{4} H \left( r
 \right)  ^{-{a}^{2}+2}{n}^{2}-4\,\Lambda\, R \left( t
 \right) ^{2\,{\frac {2\,b+3\,a}{a}}} H \left( r
 \right) ^{-{a}^{2}-ab+3} V \left( r \right)  
^{3}{r}^{4}-2\, R \left( t   \right) ^{4} H
 \left( r \right) ^{-2\,{a}^{2}}{\alpha}^{2} h^2V \left( 
r \right) ^{2}\nn\label{E3}.\\
&&
\end{eqnarray}
The numerator of ${\cal G}_{\phi\phi}$ is a combination of two terms as $\bar {\cal G}_{\phi\phi}=\cos ^2\theta \bar {\cal G}^{(1)}_{\phi\phi}+\bar {\cal G}^{(2)}_{\phi\phi}$ where
\begin{eqnarray}
\bar{\cal G}^{(2)}_{\phi\phi}&=&-6R \left( t \right) ^{4} H \left( r
 \right)  ^{- \left( a-1 \right)  \left( a+1 \right) } V \left( r 
 \right) ^{4}{r}^{5}
{\frac {d}{dr}}H \left( r \right)  -3R \left( t\right) ^{4}
 H \left( r  \right) ^{- \left( a-1 \right)  \left( a+1
 \right) }V \left( r   \right) ^{4}{r}^{6} {\frac {d^{2}}{d{r}^{2}}}H \left( r \right) 
  \nn\\&-&2R
 \left( t \right)  ^{4}{h}^{2} H \left( r 
 \right) ^{-2{a}^{2}}{r}^{2} V \left( r  \right) ^{4}
{\alpha}^{2}+3R \left( t  \right) ^{4} H
 \left( r \right)  ^{-{a}^{2}}V \left( r  \right) ^{4}{r}^{6} \left({\frac {d}{dr}}H \left( 
r \right) \right)  ^{2} \nn\\
&-&3R \left( t \right) ^{4}  H \left( r
 \right)  ^{-{a}^{2}+2}{n}^{2}  V \left( r \right) 
 ^{2}{r}^{2}+18R \left( t  \right) ^{4}
 H \left( r \right)  ^{-{a}^{2}+5}V \left( r  \right) 
^{5}{r}^{6} \left( {\frac {d}{dt
}}R \left( t \right)  \right) ^{2} \nn\\&-&6R \left( t  \right) ^{4} H
 \left( r  \right) ^{-{a}^{2}+2} V \left( r 
 \right) ^{3}  {r}^{5
}{\frac {d}{dr}}V \left( r\right) -3R \left( t \right)  ^{4} H \left( r
 \right) ^{-{a}^{2}+2} V \left( r \right)  ^{3}{r}
^{6} {\frac {d^{2}}{d{r}^{2}}}V
 \left( r \right)   \nn\\&+&3R \left( t   \right) ^{4}  H \left( r
 \right)  ^{-{a}^{2}+2}  V \left( r \right)  ^{2
} {r}^{6}\left( {\frac {d}{dr}}V \left( r \right)  \right) ^{2}-4
 R \left( t  \right) ^{2{\frac {3a+2b}{a}}}
 H \left( r \right)  ^{-{a}^{2}-ab+3}\Lambda V
 \left( r  \right) ^{5}{r}^{6}\nn\\&+&6R \left( t
 \right) ^{5}  H \left( r \right) ^{-{a}^{2}+5} 
V \left( r \right) ^{5}{r}^{6}{\frac {d^{2}}{d{t}^{2}}}R
 \left( t \right),\label{E7}
\end{eqnarray}

and $\bar {\cal G}^{(2)}_{\phi\phi}$ is given by

\begin{eqnarray}
\bar {\cal G}^{(1)}_{\phi\phi}&=&
3R \left( t\right) ^{4}  H \left( r  
 \right) ^{- \left( a-1 \right)  \left( a+1 \right) } { V \left( r
 \right) ^{4}{r}^{6}\frac {d
^{2}}{d{r}^{2}}}H \left( r \right) +6 R \left( t \right) ^
{4}  H \left( r \right) ^{- \left( a-1 \right)  \left( 
a+1 \right) }  
 V \left( r \right) ^{4}{r}^{5}{\frac {d}{dr}}H \left( r \right)\nn\\&-&3 R \left( t
 \right) ^{4}  H \left( r \right) ^{-{a}^{2}+2
} {n}^{2}{r}^{4}\left( {\frac {d}{dr}}V \left( r \right)  \right) ^{2}
-6R \left( t \right) ^{5} H \left( r
 \right) ^{-{a}^{2}+5} V \left( r\right) ^{5
}{r}^{6}{\frac {d^{2}}{d{t}^{2}}}R \left( t \right) \nn\\&+&6R
 \left( t \right)  ^{4} H \left( r \right)  ^{-
{a}^{2}+2}  V \left( r \right)  ^{3} {r}^{5} {\frac {d}{d
r}}V \left( r \right) +4 R \left( t 
 \right) ^{2\,{\frac {3\,a+2\,b}{a}}} H \left( r 
 \right) ^{-{a}^{2}-ab+3}\Lambda\, V \left( r \right)  
^{5}{r}^{6}\nn\\&+&3R \left( t \right) ^{4} H
 \left( r \right) ^{-{a}^{2}+2}  V \left( r \right)  ^{3
}{r}^{6} {\frac {d^{2}}{d{r}^{2
}}}V \left( r  \right) -3 R \left( t  \right) ^{4} H \left( r
 \right) ^{-{a}^{2}+2} V \left( r \right) ^{2
} {r}^{6}\left( {\frac {d}{dr}}V \left( r \right)  \right) ^{2}\nn\\&-&18
 R \left( t  \right) ^{4} H \left( r
 \right) ^{-{a}^{2}+5} V \left( r  \right) ^{5}{r}^{6}\left( {\frac {d}{dt}}R \left( t \right) 
 \right) ^{2} +3R \left( t\right) ^{4}  H \left( r 
 \right) ^{-{a}^{2}+2}{n}^{2} V \left( r \right) ^{2}{
r}^{2}\nn\\&-&3R \left( t \right) ^{4} H \left( r
 \right) ^{-{a}^{2}} V \left( r  \right) ^{4}{r}^{6}\left( {\frac {d}{dr}}H \left( r
 \right)  \right) ^{2} -
3R \left( t \right) ^{4}H \left( r \right)  ^{- \left( a-1 \right)  \left( a+1 \right) }   V \left( r
 \right)  ^{2}{n}^{2}{r}^{4}{\frac {d
^{2}}{d{r}^{2}}}H \left( r  \right)\nn\\&-&6R \left( t \right) 
 ^{4}H \left( r \right) ^{- \left( a-1
 \right)  \left( a+1 \right) } V \left( r \right) ^{2}{n}^{2}{r}^{
3} {\frac {d}{dr}}H \left( r
 \right)   +2 R \left( t \right) ^{4}{h}^{2} H \left( 
r \right) ^{-2\,{a}^{2}}{r}^{2} V \left( r \right) 
 ^{4}{\alpha}^{2}\nn\\&+&3 R \left( t  \right) ^{4}
 H \left( r \right)  ^{-{a}^{2}}  V \left( r  \right) ^{
2}{n}^{2}{r}^{4}\left( {\frac {d}{dr}}
H \left( r \right)  \right) ^{2}+3 R \left( t \right) ^{4}  H
 \left( r \right)  ^{-{a}^{2}+2} V \left( r \right) {n}^{2}{r}^{4}{\frac {d^{2}}{d{r}^{2
}}}V \left( r \right)  \nn\\&+&18 R \left( t \right) ^{4} 
H \left( r \right) 
 ^{-{a}^{2}+5} V \left( r \right) ^{3} {n}^{2}{r}^{4}\left( 
{\frac {d}{dt}}R \left( t \right)  \right) ^{2}+6R \left( t \right)^{4} H \left( r \right) 
 ^{-{a}^{2}+2}V \left( r \right)  {n}^{2}{r}^{3} {\frac {d}{dr}}V
 \left( r \right)\nn\\&+&6R \left( t
 \right) ^{5} H \left( r \right) ^{-{a}^{2}+5
}   
V \left( r  \right) ^{3}{n}^{2}{r}^{4}{\frac {d^{2}}{d{t}^{2}}}R \left( t \right)-2{h}^{2} H
 \left( r  \right) ^{-2{a}^{2}} R \left( t \right) 
 ^{4} V \left( r   \right) ^{2}{\alpha}^{2}{n}^{
2}\nn\\&-&4\Lambda R \left( t  \right) ^{2{\frac {3a+2
b}{a}}} H \left( r  \right) ^{-{a}^{2}-ab+3} V
 \left( r \right) ^{3}{n}^{2}{r}^{4}+3R \left( t
 \right) ^{4} H \left( r  \right) ^{-{a}^{2}+2
}{n}^{4}\label{E6}.
\end{eqnarray}

The expressions for $\bar {\cal G}_{\psi\psi}$ and the only non-diagonal component of Einstein's equation $\bar {\cal G}_{\psi\phi}$ are lengthy and so we don't mention them here. We call these to equations (\ref{E7}) and (\ref{E8}) respectively in what follows. 
\begin{eqnarray}
\bar {\cal G}_{\psi\psi}&=&6R \left( t \right) ^{5}  H \left( r 
 \right) ^{-{a}^{2}+5} V \left( r \right) ^{3}{r}^{4} {\frac {d^{2}}{d{t}^{2}}}R \left( t
 \right)+18
R \left( t \right) ^{4} H \left( r 
 \right) ^{-{a}^{2}+5}V \left( r \right) ^{3} {r}^{4}\left( 
{\frac {d}{dt}}R \left( t \right)  \right) ^{2}\nn\\&+&3R
 \left( t \right) ^{4} H \left( r  \right) ^{-
{a}^{2}+2}  V \left( r \right) {r}^{4} {\frac {d^{2}}{d{r}^{2}}}V \left( r 
 \right)-3R \left( t 
 \right) ^{4}  H \left( r \right) ^{-{a}^{2}+1} 
 V \left( 
r \right) ^{2}{r}^{4}{\frac {d^{2}}{d{r}^{2}}}H \left( r \right) \nn\\ &-&3R \left( t \right) 
^{4}  H \left( r \right) ^{-{a}^{2}+2} {r}^{4}\left( {\frac {d
}{dr}}V \left( r \right)  \right) ^{2}+3 R \left( t
 \right)  ^{4} H \left( r  \right) ^{-{a}^{2}}
  V
 \left( r \right) ^{2}{r}^{4}\left( {\frac {d}{dr}}H \left( r \right)  \right) ^{2}\nn\\&+&6R \left( t 
 \right) ^{4}  H \left( r \right)  ^{-{a}^{2}+2}V
 \left( r \right)    {
r}^{3}{\frac {d}{dr}}V \left( r  \right)-6R \left( t \right) ^{4} H \left( r
 \right) ^{-{a}^{2}+1}   V \left( r\right) ^{2}{r}^{3}{\frac {d}{dr}}H \left( r
 \right)+3
 R \left( t  \right) ^{4} H \left( r \right) 
 ^{-{a}^{2}+2}{n}^{2}\nn \\ &-&4\Lambda R \left( t 
 \right) ^{2\,{\frac {3\,a+2\,b}{a}}}H \left( r 
 \right) ^{-{a}^{2}-ab+3}  V \left( r \right) ^{3}{r}^{
4}-2 R \left( t \right) ^{4} H \left( r
 \right) ^{-2\,{a}^{2}}{\alpha}^{2}{h}^{2} V \left( r
 \right) ^{2}\label{E4},
\end{eqnarray}
\begin{eqnarray}
\bar {\cal G}_{\psi\phi}&=&-3 R \left( t \right)  ^{4} H \left( r
 \right) ^{-{a}^{2}+2}{n}^{2}-6R \left( t 
 \right) ^{5}  H \left( r \right) ^{-{a}^{2}+5} {r}^{4} V \left( 
r \right) ^{3}
{\frac {d^{2}}{d{t}^{2}}}R \left( t \right)  \nn\\
&-&18R \left( t 
 \right) ^{4}  H \left( r \right) ^{-{a}^{2}+5} 
V \left( r  \right) ^{3} {r}^{4}\left( {\frac {d}{dt}}R \left( t
 \right)  \right) ^{2}-3R \left( t \right) ^
{4} H \left( r \right) ^{-{a}^{2}+2}  V \left( r \right) {r}^{4}{\frac {d^
{2}}{d{r}^{2}}}V \left( r   \right)\nn\\&+
&3R \left( t  \right) ^{4} H \left( r 
 \right) ^{- \left( a-1 \right)  \left( a+1 \right) }    V \left( r
 \right)  ^{2}{r}^{4}{\frac {d
^{2}}{d{r}^{2}}}H \left( r \right) +3R \left( t  \right) ^
{4} H \left( r \right)^{-{a}^{2}+2} {r}^{4}\left( {\frac {d}
{dr}}V \left( r \right)  \right) ^{2}\nn\\
&-&3R \left( t
 \right) ^{4} H \left( r \right)  ^{-{a}^{2}}
 V
 \left( r \right) ^{2}{r}^{4}\left( {\frac {d}{dr}}H \left( r \right)  \right) ^{2} -6 R \left( t 
 \right) ^{4} H \left( r  \right) ^{-{a}^{2}+2}V
 \left( r \right)  {
r}^{3}{\frac {d}{dr}}V \left( r \right) \nn\\
&+&6 R \left( t\right) ^{4} H \left( r
 \right)  ^{- \left( a-1 \right)  \left( a+1 \right) } V \left( r \right) 
 ^{2}{r}^{3}
{\frac {d}{dr}}H \left( r  \right)  +4\Lambda R \left( t \right) ^
{2\,{\frac {3\,a+2\,b}{a}}}  H \left( r \right) ^{-{a}^
{2}-ab+3}  V \left( r \right) ^{3}{r}^{4}\nn\\
&+&2 R
 \left( t\right) ^{4} H \left( r\right) ^{-
2\,{a}^{2}}{\alpha}^{2}{h}^{2} V \left( r \right) ^{2}.\label{E5}
\end{eqnarray}
Moreover,  the equation of motion for the dilaton (\ref{eqphi}) becomes
\begin{eqnarray}
&&36V \left( r \right)  R \left( t  \right) ^{4}
 H \left( r \right)  ^{-{a}^{2}+3}{r}^{4} \left( {
\frac {d}{dt}}R \left( t \right)  \right) ^{2}+12V \left( r \right) 
 R \left( t \right) ^{5} H \left( r 
 \right) ^{-{a}^{2}+3}{r}^{4}{\frac {d^{2}}{d{t}^{2}}}R \left( t
 \right) \nn\\&+&3\, R \left( t \right) ^{4}  H \left( 
r   \right) ^{-{a}^{2}-1}{r}^{4}{a}^{2}{\frac {d^{2}}{d{r}^{2}}
}H \left( r \right) -3\, R \left( t \right) ^{4}
 H \left( r \right) ^{-{a}^{2}-2}{r}^{4}{a}^{2}\left(
 {\frac {d}{dr}}H \left( r \right) \right) ^{2}\nn\\&+&6R
 \left( t\right) ^{4} H \left( r\right) ^{-
{a}^{2}-1}{a}^{2} {r
}^{3}{\frac {d}{dr}}H \left( r \right) +4b R \left( t  \right) ^{2{\frac {3a+2b}{
a}}}  H \left( r \right)   ^{-{a}^{2}-ab+1}\Lambda\,V
 \left( r \right) {r}^{4}a+2\,{a}^{2}{\alpha}^{2}{h}^{2} H
 \left( r \right)  ^{-2\,{a}^{2}-2} R \left( t \right) 
^{4}\nn\\&=&0\label{E8}.
\end{eqnarray}
From $\bar {\cal G}_{tt}=0$ in (\ref{E1}), we find $\alpha^2 h^2$ in terms of other unknown quantities $\Lambda$, $b$, $H$ and $R$ and substitute in all other equations (\ref{E2})-(\ref{E8}) (setp 1). We then consider equation (\ref{E3}) that was obtained in step 1 and solve for $\Lambda$ in terms of other unknwon quantities $b$, $H$ and $R$. In step 2, we substitute for $\Lambda$ in equations (\ref{E4}),(\ref{E5}),(\ref{E6}),(\ref{E7}) and (\ref{E8}) that were obtained in step 1. The equations (\ref{E4}),(\ref{E5}),(\ref{E6}) and (\ref{E7}) in step 2 are satisfied upon substituting $V(r)=1+\frac{n}{r}$. It turns out the equation (\ref{E8}) in step 2 is a differential equation only for $R(t)$ with the solution $R(t)=(R_0 t+R_1)^m$ where $m=-\frac{a^2-2ab-2}{4(ab+1+b/a)}$ and $R_0$ and $R_1$ are two constants. However this solution leads to a time dependent or a vanishing cosmological constant unless we choose $ab=-2$. So, setting $ab=-2$, we find the differential equation for the metric function $R(t)$ as
\be
\left( {\frac {dR}{dt}}\right) ^{2}({a}^{3}-2a-\frac{8}{a})-R  \left( {\frac {d^{2}R}{d{t}^{2}}} \right) 
({a}^{3}+2a)=0,
\ee 
with the solutions $R(t)=(R_0 t+R_1)^{\frac{a^2}{4}}$. Without loss of generality, we choose $R_1=0$.
Using the solutions for $R(t)$ in equation (\ref{E2}) in step 1, we find the differential equation for the metric function $H(r)$,
\be
\left( {\frac {dH}{dr}} \right) ^{2} 
%\left(
 %H(r)
 %\right) ^{-{a}^{2}}
{a}^{2}r+2r\, \left( {\frac {d^{2
}H}{d{r}^{2}}}   \right)  H \left( r \right) 
%^{-{a}^{2}+1}
+4\, \left( {\frac {dH}{dr}}  
 \right)   H \left( r \right) =0,
\ee
with the solution as $H(r)=(h_0+\frac{h}{r})^{\frac{2}{2+a^2}}$. We choose $h_0=1$.\vspace{1cm}
\newline
{\Large Acknowledgments}

This work was supported by the Natural Sciences and Engineering Research
Council of Canada.

\end{document}